\documentclass{PoS}

\usepackage[utf8]{inputenc}
\usepackage[shortlabels]{enumitem}
\usepackage{amsmath}
\usepackage{cite}

\title{The Large-$n_f$ Limit of the Four-Loop Splitting Functions in QCD}

\ShortTitle{The Large-$n_f$ Limit of the Four-Loop Splitting Functions in QCD}

\author{\speaker{Joshua Davies}\\
        Institut f\"ur Theoretische Teilchenphysik, Karlsruhe Institute of Technology (KIT)\\
        76128 Karlsruhe, Germany\\
        E-mail: \email{joshua.davies@kit.edu}}

\abstract{These proceedings describe a computation of the large-$\nf$ terms contributing to the QCD splitting functions at the fourth order in the strong coupling constant $\als$. Using the \texttt{FORCER} package for the reduction of four-loop two-point Feynman integrals, Mellin moments of the four-loop splitting functions have been computed. These moments are used to derive analytic Mellin-space expressions, by forming and finding solutions to systems of Diophantine equations. Expressions for the terms proportional to $\nfthr$ of the flavour singlet, and to $\nftwo$ of the flavour non-singlet splitting functions have been determined.}

\FullConference{XXV International Workshop on Deep-Inelastic Scattering and Related Subjects\\
		3-7 April 2017\\
		University of Birmingham, UK}

\newcommand{\als}{\alpha_s}
\newcommand{\as}{a_s}

\newcommand{\cf}{C_F}
\newcommand{\cftwo}{C^{\,2}_F}

\newcommand{\ca}{C_A}

\newcommand{\nf}{n_f}
\newcommand{\nftwo}{n^{\,2}_f}
\newcommand{\nfthr}{n^{\,3}_f}

\newcommand{\mufsq}{\mu_f^{\,2}}

\newcommand{\Pqq}{P_{\rm qq}}
\newcommand{\Pqg}{P_{\rm qg}}
\newcommand{\Pgq}{P_{\rm gq}}
\newcommand{\Pgg}{P_{\rm gg}}
\newcommand{\Pab}{P_{\rm ab}}
\newcommand{\Pnsp}{P_{\rm ns}^+}
\newcommand{\Pnsm}{P_{\rm ns}^-}
\newcommand{\Pnspm}{P_{\rm ns}^\pm}
\newcommand{\Pv}{P_{\rm v}}
\newcommand{\Pqqn}[1]{P_{\rm qq}^{(#1)}}

\newcommand{\Pgqn}[1]{P_{\rm gq}^{(#1)}}
\newcommand{\Pggn}[1]{P_{\rm gg}^{(#1)}}

\newcommand{\Pnspn}[1]{P_{\rm ns}^{(#1),+}}
\newcommand{\Pnsmn}[1]{P_{\rm ns}^{(#1),-}}
\newcommand{\Pnspmn}[1]{P_{\rm ns}^{(#1),\pm}}
\newcommand{\Pvn}[1]{P_{\rm v}^{(#1)}}

\newcommand{\qa}{q_{\rm a}}
\newcommand{\qb}{q_{\rm b}}
\newcommand{\qba}{\bar{q}_{\rm a}}
\newcommand{\qbb}{\bar{q}_{\rm b}}
\newcommand{\qs}{q_{\rm s}}
\newcommand{\qabpm}{q_{\rm ab}^\pm}
\newcommand{\qv}{q_{\rm v}}
\newcommand{\fa}{f_{\rm a}}
\newcommand{\fb}{f_{\rm b}}

\newcommand{\FORCER}{\texttt{FORCER}}
\newcommand{\axb}{\texttt{axb}}

\begin{document}

\section{Introduction}
In recent years, the next-to-next-to-leading order (NNLO) perturbative QCD corrections have been completed for many processes. For a consistent treatment, such processes with protons in their initial state require parton distribution functions which are evolved at the same order. The three-loop splitting functions governing their evolution have been known for some time \cite{Moch:2004pa,Vogt:2004mw}.

For a few processes, the N$^3$LO QCD corrections are known \cite{Anzai:2015wma,Anastasiou:2016cez,Vermaseren:2005qc,Moch:2008fj}. An analysis at this order requires, in principle, the four-loop splitting functions. The first results for low-$N$ Mellin moments of these contributions were presented in \cite{Baikov:2006ai,Velizhanin:2011es,Velizhanin:2014fua,Baikov:2015tea}; the flavour non-singlet quark$+$anti-quark splitting function $\Pnspn{3}$ was computed at $N=2,4$ and its quark$-$anti-quark counterpart $\Pnsmn{3}$ at only $N=3$.

The recently developed \FORCER{} package \cite{Ruijl:2017cxj} for the parametric reduction of four-loop self-energy integrals has allowed significant progress to be made; the non-singlet splitting functions are now known for values up to $N=16$ \cite{Moch:2017uml} and the singlet splitting functions for values up to $N=4$ \cite{Ruijl:2016pkm}.

The subsets of Feynman diagrams contributing to the leading powers of $\nf$ (the number of massless quark flavours) are much easier to compute. The most difficult four-loop topologies do not contribute, and \FORCER{} can provide Mellin moments for much higher values of $N$. The number of moments computed for these terms is sufficiently large that when combined with suitable functional ansatzes, based on the structure of lower-order splitting function contributions, analytic expressions can be derived for their dependence on $N$ using lattice basis reduction techniques \cite{LLL,AXB}. See \cite{Velizhanin:2012nm,Moch:2014sna} for earlier works using these methods.

We now discuss how these calculations are performed and consider how the results compare to existing literature. For a more complete discussion and the results of the calculations, which are not presented in these proceedings, the reader is referred to Ref.~\cite{Davies:2016jie}.

\section{Calculation}
\label{sec:calculation}
In Mellin space, the scale evolution of the parton distribution functions is given by
\begin{equation}
	\frac{d}{d\ln \mufsq} \: \fa = \Pab \: \fb,
\end{equation}
where $\fa = q_u, \bar{q}_u, \ldots, g$. This system of $2\nf+1$ coupled equations can be decomposed into a $2\times2$ system for the coupled evolution of the flavour singlet $\left( \qs = \sum_{i=1}^{\nf} (q_i+\bar{q}_i) \right)$ and gluon $(g)$ distributions,
\begin{equation}
	\frac{\mbox{d}}{\mbox{d}\ln\mufsq}
	\left(
	\begin{array}{c}
		\qs \\
		g \\
	\end{array}
	\right)
	=
	\left(
	\begin{array}{cc}
		\Pqq & \Pqg \\
		\Pgq & \Pgg \\
	\end{array}
	\right)
	\left(
	\begin{array}{c}
		\qs \\
		g \\
	\end{array}
	\right),
\label{eqn:singletevo}
\end{equation}
and $2\nf-1$ equations for the evolution of the flavour non-singlet $\left( \qabpm = \left(\qa\!\pm\!\qba\right)\!-\!\left(\qb\!\pm\!\qbb\right) \right)$ and valence $\left( \qv = \sum_{a=1}^{\nf}(\qa-\qba) \right)$ distributions,
\begin{equation}
	\frac{\mbox{d}}{\mbox{d}\ln\mufsq} \qabpm = \Pnspm \: \qabpm, \qquad
	\frac{\mbox{d}}{\mbox{d}\ln\mufsq} \qv = \Pv \: \qv.
\label{eqn:nsvalenceevo}
\end{equation}
It is the seven splitting functions appearing in Eq. (\ref{eqn:singletevo}) and Eq. (\ref{eqn:nsvalenceevo}) that we are concerned with here. We compute them in the context of deep inelastic scattering (DIS) in the style of Ref.~\cite{Larin:1996wd}, to which the reader is referred for further details. A brief description is given here.

For inclusive DIS, one can use the optical theorem to map the partonic cross sections to forward amplitudes,
\begin{equation}
	\mbox{probe}(q) + \mbox{parton}(p) \rightarrow \mbox{probe}(q) + \mbox{parton}(p),
\end{equation}
where $p^2 = 0$ and $q^2 = -Q^2 < 0$. The forward amplitudes are projected onto powers of the parton momentum ($p$) and a dispersion relation then yields, as the coefficients of $\left( 2p\cdot q / Q^2 \right){}^N$, even- or odd-$N$ Mellin moments of the partonic cross sections (depending on which is being considered). It is this projection which yields the four-loop self-energy-type Feynman integrals which can be efficiently reduced to master integrals by \FORCER{}.

Working in $D=4-2\epsilon$ dimensions, the partonic cross sections contain poles in the parameter $\epsilon$. The $n$-loop contributions to the splitting functions can be determined from the coefficient of $\as^n/\epsilon$ of the partonic cross sections (where $\as = \als/4\pi$). We obtain even-$N$ moments for the singlet splitting functions $\Pab$ and the non-singlet splitting function $\Pnsp$ and odd-$N$ moments for the non-singlet splitting function $\Pnsm$ and the valence splitting function $\Pv$.

\section{Determining Analytic Expressions}
Section \ref{sec:calculation} briefly describes how one can obtain Mellin moments of the splitting functions from the calculation of partonic cross sections in DIS. Here we outline the method by which we determine analytic expressions for the $N$ dependence of the splitting functions, from knowledge of a few of their Mellin moments.

To the order at which they are known (in both fixed-order results \cite{Moch:2004pa,Vogt:2004mw} and all-order resummations of certain leading-$\nf$ terms \cite{Gracey:1994nn,Gracey:1996ad,Bennett:1998sr}), the splitting functions can be expressed in terms of products of simple denominators $D_i^p = (N+i)^{-p}$ and harmonic sums, defined recursively as
\begin{equation}
	S_{\pm m}(N) = \sum_{i=1}^{N} \frac{(\pm1)^i}{i^m}, \qquad
	S_{\pm m_1,m_2,\ldots,m_l}(N) = \sum_{i=1}^{N} \frac{(\pm1)^i}{i^{m_1}} S_{m_2,\ldots,m_l}(i).
\end{equation}
The \emph{weight} $w$ of a harmonic sum is defined as $\sum_{i=1}^l |m_i|$, and the \emph{overall weight} of a harmonic sum and $D_i^p$ combination is defined to be $(w+p)$. The $\as^n$ contributions to splitting functions contain terms with overall weight up to $(2n-1)$. Terms proportional to Riemann-Zeta values $\zeta_m$ have a maximal overall weight reduced by $m$.

With these definitions, a contribution to a splitting function $\as^{n+1} P^{(n)}$ may be assumed to have the following structure,
\begin{equation}
	P^{(n)}(N) = \sum_{w=0}^{2n+1} c_{00w} \: Sw(N)
		+ \sum_i \sum_{p=0}^{2n+1} \:\: \sum_{w=0}^{2n+1-p} c_{ipw} \: D_i^p \: Sw(N),
	\label{eqn:ansatz}
\end{equation}
where $Sw(N)$ denotes the set of harmonic sums of weight $w$ (excluding those with an index ``-1'', which do not appear in any known splitting function) and we define $S_0(N) = 1$. It should be noted that at three loops, the renormalized DIS partonic cross sections already contain structures which cannot be written in this way. Nevertheless, we assume this structure to be valid at least for the large-$\nf$ terms of the four-loop splitting functions, which we consider here.

Given a sufficient number of Mellin moments of $P^{(n)}$, one can of course determine all coefficients $c_{00w}$ and $c_{ipw}$ of Eq. (\ref{eqn:ansatz}). In general, however, such a structure contains far too many coefficients to be able to compute a sufficient number of Mellin moments for their determination. We proceed by noting that, up to predictable powers of $(1/3)$, $c_{00w}$ and $c_{ipw}$ are all \emph{integers}. The system of equations formed by equating Eq.~(\ref{eqn:ansatz}) with the Mellin moments for different values of $N$ is a \emph{Diophantine} system and can be solved with fewer equations than unknown coefficients. For this we use the routine \axb{} of \cite{calc}.

For the singlet splitting functions, we determine the leading-$\nf$ colour factors $\cf\nfthr$ of $\Pqq^{(3)}$ and $\Pgq^{(3)}$, and $\cf\nfthr$ and $\ca\nfthr$ of $\Pqg^{(3)}$ and $\Pgg^{(3)}$. The above considerations suffice to determine solutions to the systems of equations for all but the $\ca\nfthr$ terms of $\Pqg^{(3)}$. For this contribution we must additionally assume that:
\vspace{-\topsep}
\begin{enumerate}[(i)]
	\setlength\itemsep{0mm}
	\item in the large-$N$ limit, the only constants which may appear are Riemann-Zeta values,
	\item the coefficients of $S_{1,2}$ are equal to those of $S_{2,1}$, up to a sign change.
\end{enumerate}
\vspace{-\topsep}
These assumptions are based on observations of the lower-order contributions to $\Pqg$. In this case, we have an ansatz of 117 unknown integer coefficients which are determined by \axb{} using Mellin moments at $N=2,4,\ldots,44$ with $N=46$ providing a check of the result.

The $\nfthr$ terms of the non-singlet splitting functions are already known to all orders in $\as$ \cite{Gracey:1994nn}. Here, we determine analytic expressions for the terms proportional to $\cftwo\nftwo$ and $\ca\cf\nftwo$. The subsets of diagrams which provide these colour factors are significantly harder to compute with \FORCER{} than those of the singlet case described above. We can not compute so many moments, and must appeal to additional structure in order to solve the systems of equations. The crucial observation is that if one writes $\Pnspmn{3}$ in the following way,
\begin{equation}
	\Pnspmn{3}\Big|_{\nftwo} =\:\: 2\cftwo A + (\ca-2\cf)B^\pm =\: 2\cftwo (A - B^\pm) + \ca\cf B^\pm,
	\label{eqn:Pnsdecomp}
\end{equation}
the function $A$ is \emph{common to both} $\Pnspmn{3}$. This means that one can use both the even- and odd-$N$ moments of $\Pnspn{3}$ and $\Pnsmn{3}$ to determine it, thus obtaining a sufficient number of moments without the value of $N$ becoming prohibitively high. The linear combinations $(A-B^\pm)$ of the second part of Eq. (\ref{eqn:Pnsdecomp}) can be determined by computing just the (easier) $\cftwo\nftwo$ diagrams. Even so, in order to obtain solutions for $(A-B^\pm)$ we enforce, in addition to (i) and (ii) (without the sign change) above, that
\vspace{-\topsep}
\begin{enumerate}[(i)]
	\setlength\itemsep{0mm}
	\setcounter{enumi}{2}
	\item $\Pnspn{3}$ and $\Pnsmn{3}$ behave as $\ln N$ in the large-$N$ limit \cite{Korchemsky:1988si,Albino:2000cp}
\end{enumerate}
\vspace{-\topsep}
to eliminate more coefficients from the ansatz. With 115 coefficients remaining, $(A-B^+)$ is determined with even moments $N=2,\ldots,40$ and $(A-B^-)$ with odd $N=3,\ldots,37$. In each case one additional moment provides verification. To determine $A$ we must \emph{further} assume that
\vspace{-\topsep}
\begin{enumerate}[(i)]
	\setlength\itemsep{0mm}
	\setcounter{enumi}{3}
	\item only positive-index harmonic sums appear,
	\item no ``many-index'' sums appear. We remove $S_{1,1,2}$, $S_{1,1,1,2}$ etc. from the ansatz.
\end{enumerate}
\vspace{-\topsep}
With these assumptions, $A$ can be determined from an ansatz of 55 unknown coefficients using moments $N=2,3,\ldots,17$ with $N=18,19,\ldots,22$ providing verification.

We now list existing literature against which we can verify our new results:
\vspace{-\topsep}
\begin{itemize}
	\setlength\itemsep{0mm}
	\item Refs.~\cite{Gracey:1996ad,Bennett:1998sr} give the $\nfthr$ terms for two linear combinations of $\Pqqn{3}$, $\Pgqn{3}$ and $\Pgqn{3}$, $\Pggn{3}$.
	\item The function $A$ of Eq.~(\ref{eqn:Pnsdecomp}) yields, in the large-$N$ limit, the large-$n_c$ limit of the cusp anomalous dimension. It is in agreement with Refs.~\cite{Henn:2016men,Grozin:2016ydd}.
	\item Ref.~\cite{Dokshitzer:2005bf} provides a prediction for the $\frac{\ln N}{N}$ coefficients of $\Pnspmn{3}$ in the large-$N$ limit, in terms of lower-order coefficients of $\ln N$. This prediction is verified here at $\nftwo$.
	\item The highest three double logarithms in both the large- and small-$x$ limits have been predicted in Refs.~\cite{Soar:2009yh,Vogt:2012gb,smallxresum} and agree with our results.
\end{itemize}
\vspace{-\topsep}

\section{Conclusions}
The development of \FORCER{} has allowed a much greater number of Mellin moments of the four-loop splitting functions to be computed than was previously possible. The diagrams contributing to the leading terms in the large-$\nf$ limit can be computed to sufficiently high values of $N$ that we are able to determine analytic expressions for their $N$ dependence. By choosing a suitable ansatz of basis functions (Eq. (\ref{eqn:ansatz})) and making various additional assumptions about the structure ((i)-(v) above) we are able to solve the systems of Diophantine equations for the Mellin moments using \axb{}.

We find solutions for the $\nfthr$ terms of the singlet splitting functions and terms proportional to $\nftwo$ of the non-singlet splitting functions. We have also determined, but not discussed here, the terms proportional to $\nftwo\,d^{abc}d_{abc}/n_c$ of the valence splitting function $\Pvn{3}$. Our results are presented in both Mellin-$N$ and $x$ space in Ref.~\cite{Davies:2016jie}.

For the remaining colour factors, the moment calculations become very computationally demanding. This, in addition to much larger ansatzes, means that this method cannot be used to determine analytic expressions for the full four-loop splitting functions. Numerical approximations to the remaining colour factors can be made using the available low-$N$ moments and knowledge of their behaviour in the large- and small-$x$ limits. These are presented in Ref.~\cite{Moch:2017uml}.

\end{document}